\documentclass[aps,prb,preprint]{revtex4-1}
\usepackage{graphicx,float,amssymb,subfigure,dcolumn,epstopdf}
\usepackage{multirow}
\usepackage{bm}
\begin{document}
\title{Tunning Spin Hall conductivities in GeTe by Ferroelectric Polarizations }
\author{Wenxu Zhang\footnote{Corresponding author. E-mail address: xwzhang@uestc.edu.cn}, Zhao Teng}
\affiliation{State Key Laboratory of Electronic Thin Films and Integrated Devices,
	University of Electronic Science and Technology of China, Chengdu, 610054, P. R. China}
\author{Hongbin Zhang}
\affiliation{Institute of Materials Science, Technische Universt\"at Darmstadt, Darmstadt, 64287, Germany}
\author{Jakub \v{Z}elezn\'{y}}
 \affiliation{Institute of Physics, Czech Academy of Sciences, Cukrovarnick\'{a} 10, 16000 Prague 6, Czech Republic}
\author{ Huizhong Zeng, Wanli Zhang}
\affiliation{State Key Laboratory of Electronic Thin Films and Integrated Devices,
	 	University of Electronic Science and Technology of China, Chengdu, 610054, P. R. China}
\date{\today}
\begin{abstract}
 Controlling charge-spin current conversion by electric fields is crucial in spintronic devices, which can be realized in diatom ferroelectric semiconductor GeTe where it is established that ferroelectricity can change the spin texture. We demonstrated that the spin Hall conductivity (SHC) can be further tuned by ferroelectricity based on the density functional theory calculations. The spin texture variation driven by the electric fields was elucidated from the symmetry point of view, highlighting the interlocked spin and orbital degrees of freedom. We observed that the origin of SHC can be attributed to the Rashba effect and the intrinsic spin orbit coupling. The magnitude of one component of SHC $\sigma^z_{xy}$  can reach as large as 100 $\hbar/e(\Omega cm)^{-1}$ in the vicinity of the band edge, which is promising for engineering spintronic devices. Our work on tunable spin transport properties via the ferroelectric polarization brings novel assets into the field of spintronics.
\end{abstract}
\maketitle
\section{introduction}
Spintronics is a multidisciplinary subject where the coupling of spin and orbital degrees of freedom of electrons plays an important role. The most important consequence of this coupling is the conversion of charge current into spin current, which is called spin Hall effects (SHE). This interconversion has be found in many different materials, such as heavy metals\cite{Tanaka2008}, semiconductors\cite{Kato2004}, and interfaces between oxides, for example LAO/STO\cite{Pai2018}. Manipulation of the charge-spin conversion is one of the keys to successful applications spintronic devices\cite{Datta1990,Choi2018}. The controllable spin-orbital coupling leading to a gateable electric signal in LAO/STO was recently shown by several groups, as reviewed by Han et al\cite{Han2018}. The underlying mechanism is the change of the electron density which changes the occupation of different orbitals \cite{Lesne2016}.

\par Very recently, another material family was proposed to show electrically controllable spin-orbit coupling strength based on ferroelectricity, called ferroelectric Rashba semiconductors(FERSC). These materials are GeTe\cite{Deringer2012,Sante2012}, SnTe\cite{Plekhanov2014} and MX$_2$ \cite{Bruyer2016} etc. Among them, GeTe, a narrow gap semiconductor, received special interests due to its range of properties for modern functional materials such as thermoelectricity\cite{Wu2014}, ferroelectricity \cite{Polking2012,Kolobov2014,Kooi2016}, or electrical phase change \cite{Bruns2009} within such simple stoichiometry and the two-atom rhombohedral unit cell \cite{Boschker2017}. Both experiments and theoretical calculation show that the Rashba coupling strength in GeTe can be tuned by the electric polarization, meanwhile the spin texture is also related to the electric polarization as shown by theoretical calculations\cite{Sante2012} and spin resolved ARPES\cite{Rinaldi2018}: The clockwise winding directions in the reciprocal space of the spin vectors can be reversed to anti-clockwise. There are several other ferroelectric compounds which show the reversible spin texture controlled by electric polarization, such as HfO$_2$\cite{Tao2017},BiTeI\cite{Maas2016}, GeTe/InP superlattice\cite{Meng2017}. The controllable spin texture may lead to tuning of the spin life time, so that the proposal of spin transistor\cite{Datta1990} may be realized, as in the recently demonstrated ferromagnetic-free all-electric spin Hall transistor \cite{Choi2018}. Although the spin texture can be tuned by the polarization as clearly shown by the ARPES, it is still not shown how the spin current to charge current conversion is influenced. This conversion is the crucial step to utilize the effect in real devices.The spin to charge conversion in GeTe was observed by Rinaldi \emph{et al.}\cite{Rinaldi2016} for the first time. However, the tuning of the conversion efficiency was not reported yet.

\par In this work,  we show by \textit{ab initio} calculations, as well as by
general argument based on the effect of electric polarization on the spin texture, that the electric field can be used to tune the spin Hall conductivity (SHC) which is contributed from both the Rashba effect due to polarization and the bulk effect. The latter one is only marginally influenced, while the former effect near the band edge changed obviously under ferroelectric distortions and even the sign can be changed. It is thus possible to tune the magnitude of the SHC.

\par In the following, we first outline the theoretical background and calculation details of the electronic structure and SHC in Sec. \ref{sec:calc}. Then we analyse the electronic band structure to argue that the dependence of the spin winding directions and the electric polarization is equivalent to switching of the coordinate system in Sec. \ref{sec:bands}. We show the electric field dependence of the Rashba parameters in Sec.\ref{sec:polar}, and finally we show the polarization dependent SHC in Sec. \ref{sec:shc}.

\section{Calculation details}\label{sec:calc}
GeTe crystallizes into the noncentrosymmetric rhombohedral structure (space group R3m, No. 160) with rhombohedral  lattice constant $a=4.373$ \AA\ and the angle $\alpha$ between the axis $57.76^\circ$. Ferroelectric polarization is realized by the relative displacement of Ge and Te along the [111]-direction which is chosen as the $z$-axis in our calculations. The lattice parameters are fully relaxed when the polarization is changed. The electronic structure was calculated by the psuedo-potential plane wave method (PWscf) \cite{Giannozzi2009} within the generalized gradient approximation (GGA) parameterized by Perdew \emph{et al}.\cite{Perdew1996}. The energy cut-off was set to 30 Ry and the Brillouzin zone integration was performed on a uniform $k$-grid with $20\times20\times20$ $k$-points. The electronic structure analyse was carried out by the graphic interface VNL\cite{vnl}. The Hilbert space with the plane wave basis was projected onto the maximally localized Wannier orbital spaces spanned the valence $s$- and $p$-orbitals. Calculations of the spin Hall conductivity was performed by the code developed by one of the authors\cite{shccode}, which evaluates the Kubo formula within the linear response theory. The intrinsic SHC can be obtained with the help of Berry curvature\cite{Sinova2015}, which has the following form

\begin{equation}
\sigma^\gamma_{\alpha\beta}=\frac{e}{\hbar}\sum_n\int_{BZ}\frac{d^3\vec{k}}{(2\pi)^3}f_n(\vec{k})\Omega^\gamma_{n,\alpha\beta}(\vec{k}),
\end{equation}
where $\Omega^\gamma_{n,\alpha\beta}(\vec{k})$ is referred to as the spin Berry curvature defined as
\begin{equation}
\Omega^\gamma_{n,\alpha\beta}(\vec{k})=2i\hbar^2\sum_{m\neq n}\frac{\langle u_n(\vec{k})|\hat{J}_\alpha^\gamma|u_m(\vec{k})\rangle \langle u_n(\vec{k})|\hat{v}_\beta|u_m(\vec{k})\rangle}{(\epsilon_n(\vec{k})-\epsilon_m(\vec{k}))^2},
\end{equation}
with $\alpha,\beta,\gamma=x,y,z$, and $m,n$ being the band indices. The spin current operator is $\hat J_\alpha^\gamma=\frac{1}{2}\{{\hat{v}_\alpha,\hat{\sigma}_\gamma}\}$, where $\hat{\sigma}_\gamma$ is the spin operator component $\gamma$. The Fermi Dirac distribution function $f_n({\vec{k}})$ is the mean occupation number of state $(n,\vec{k})$ at a finite temperature T. In this work, we set $T=0$ K.  The third-order tensor $\sigma^\gamma_{\alpha\beta}$ represents the spin current $Js_\alpha^\gamma$ generated by an electric field $\vec{D}$ via $Js_\alpha^\gamma=\sigma^\gamma_{\alpha\beta} D_\beta$. The spin current is polarized in the $\gamma$ direction and flows in the $\alpha$-direction, for an electric field applied in the $\beta$-direction.

\par The integral over the $k$-space during the calculation of the SHC was sampled in the first BZ with grids of $200\times200\times200$ to ensure the convergence. The unit of SHC is $\hbar/e$($\Omega^{-1} $cm$^{-1}$).  

\section{Results and discussions}
\subsection{Electronic Structure and Ferroelectricity}\label{sec:bands}
The Kohn-Sham band structure calculated using the experimental lattice constants is shown in Fig. \ref{fig:band} where also comparison of the results with and without the SOC is made. Without the SOC, the ferroelectric distortion leads to the direct band gap of about 0.66 eV near the $L$-point as shown by the dashed blue curves. The value is only 0.05 eV higher than the experimental value reported by Park \emph{et al.}\cite{Park2009}. We noticed that even in the absence of SOC, there are humps around the $L$-point due to the electric polarization. The SOC leads to splitting ($\Delta_{SOC}$) of the bands about 0.7 eV along $\Gamma-L$ line, which indicates that the SOC field is more or less collinear in this direction. However, the Rashba like splitting are only around the $L$-point, and mostly obvious at the VBM at $L$ as shown by the solid green lines. This direction is in the direction of polarization in our coordinate system. The contributions of the SOC are mainly observed in the valence bands due to the fact that they are mostly from the Te $5p$-orbitals, while the conduction bands are from the Ge $2p$-orbitals. This is in contrast to a simple interpretation of resonant bonds, because the $5p$-states of Te is much higher in energy than the $2p$-states of Ge. The so-called ``cross gap hybridization'' was used to depict this phenomenon in the highly ionic compound\cite{Waghmare2003,Singh2013}. This anomaly is related to the enhanced Born effective charge which in turn improves the bonding with distortion.
\begin{figure}
	\caption{\label{fig:band} (color online) Comparison of the Kohn-Sham band of GeTe without ( dashed blue curves) and with SOC (solid green curves). The vertical double arrow indicates the SOC splitting of the states $\Delta_{SOC}$ along the $\Lambda$-line. The inset shows the path and the special $k$-points in the Brillouin Zone used in the calculations.}
\end{figure}

\begin{figure}
	\includegraphics[scale=0.75]{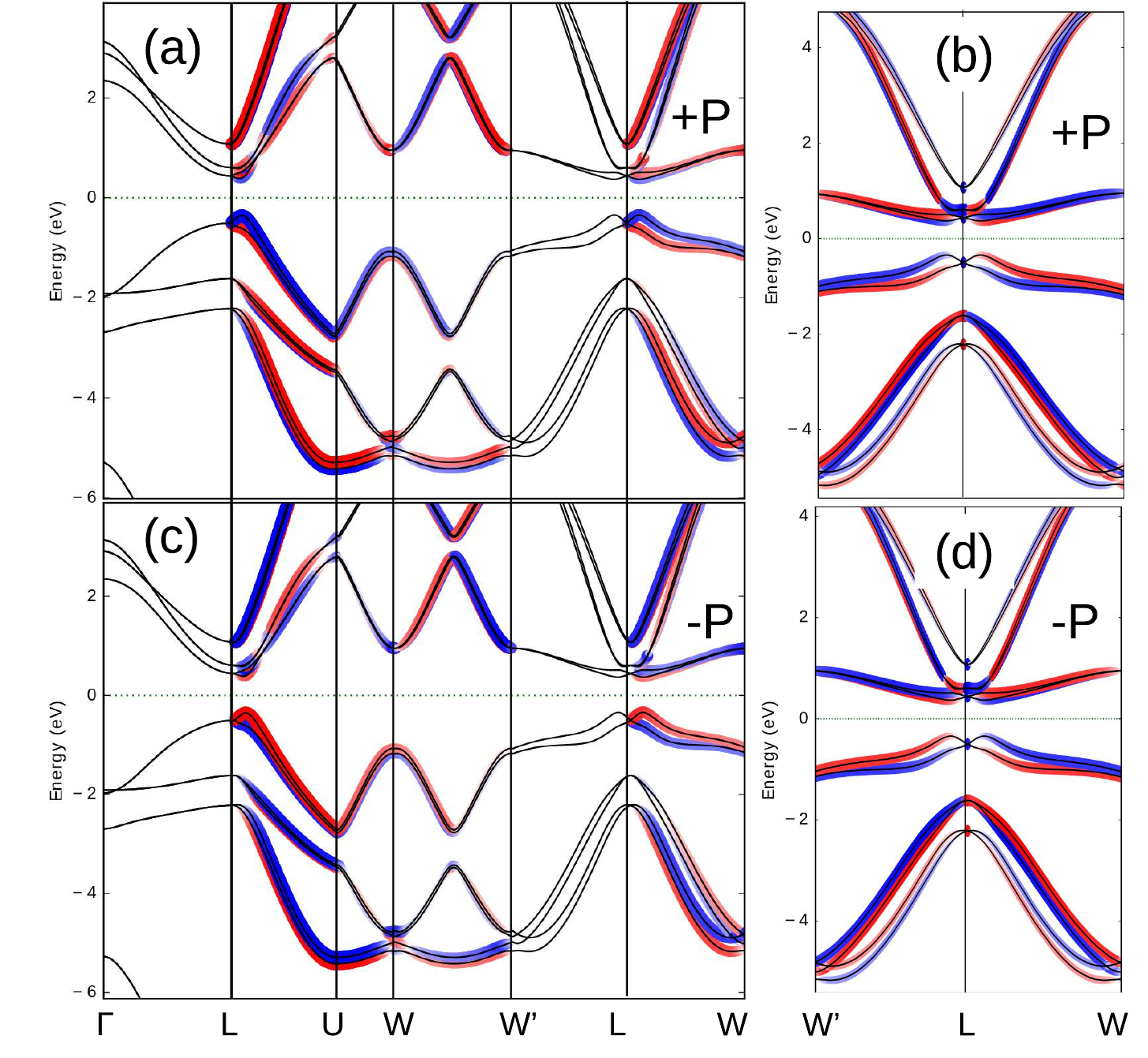}
	\caption{\label{fig:fatband} The spin component weighted band structure of GeTe with polarization of +P weighted by $x$-(a)and $z$-components(b) together with -P by $y$-(c) and $z$-components(d).}
\end{figure}
\begin{figure}
	\includegraphics[scale=0.75]{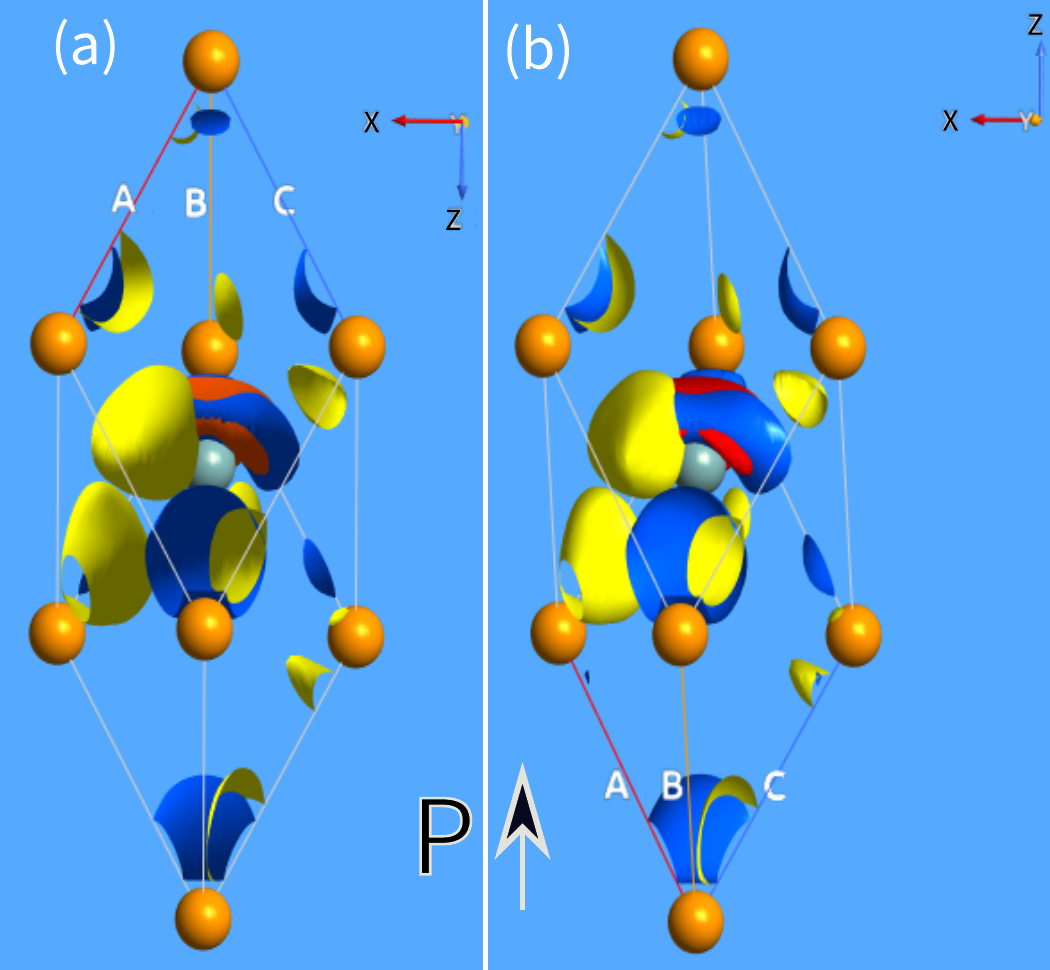}
	\caption{\label{fig:iso} (Color online) The electron isosurface of different spins with ferroelectric polarization $+P$(a) and $-P$(b). Noticed that in the two subfigures, the polarization are plotted in the same directions while the $z-$axis of coordinate system is turned upside down}
\end{figure}

The spin dependent fat band are shown in Fig. \ref{fig:fatband}.  The dispersions of the bands for reversed polarizations, namely +P and -P are the same. This is understandable because the ferroelectric displacement and spin-$z$ directions are the same. If the SOC was overlooked, the spin polarization would be independent of the crystal coordinates. In this case, the spins are unaltered upon electric polarization reversal. At the $L$-point, the spin vectors in the $X$ and $Y$ directions have different contributions when the $k$-path is along $L-U$ and $L-A$ directions as shown in Fig.\ref{fig:fatband}(a) and (c). This contributions are reversed when the polarization is reversed from $+P$ to $-P$. As a result, due to the SOC, the winding directions of the spin vectors are reversed in the $k$-space as reported in previous works \cite{Sante2012}. We noticed that there is no $x$ nor $y$-components contributions along $\Gamma-L$ coincided with the collinear SOC field in this direction as shown before. This leads to an almost rigid shift of the bands along this direction as discussed above. At the same time, we plotted the $z$-component of the spin in Fig. \ref{fig:fatband} (b) and (d). It can be seen that the different polarization  also changes the projection weight. The changing of the spin winding direction can be understood as following: In real space, the polarization reversal is equivalent to change the $z$-component of the coordinate system. In this case, the $x$-$y$ axis changes from the right-handed to the left-handed as the view points are changed from the $z$ to $-z$-axis. This effect is universal in ferroelectric materials as it is related to the changing of the coordinate system once the polarization reversal is realized only by the relative shift of the atomic positions. This effect can also be clearly seen from the electron isosurface of the different spin components as in Fig. \ref{fig:iso}. In the figure, we show the negative polarization in a cell which is rotated up-side-down, so that the atomic positions are the same as shown by (a) and (b). As observed from the figure, in this case, the spin densities of $x$, $y$ and $z$-components are the same, so that the spin direction are the same. However, the $z$-axis of the coordinates are reversed, which gives the polarization direction. In this way, the spin directions are interlocked with the ferroelectric polarization.

\subsection{Spontanous Polarization and Rashba Coupling}\label{sec:polar}
The ferroelectricity in GeTe was well explored and understood more than fifty years ago\cite{Pawley1966},\cite{Steigmeier1970}. The soft optical phonon is the cause of the displacive movement of the diatomic compound. In GeTe, ferroelectric displacement of Te along [111] shows double well energy minimums at $\tau$ = 0.03 $a_0$ away from the middle point the diagonal. The potential barrier height is about 54 meV, which is double as that of BaTiO$_3$ and half as that of PbTiO$_3$\cite{Cohen1992}. This indicates that the Curie temperature about 670 K of GeTe lies between them. The spontaneous ferroelectric polarization of 67 $\mu$C/cm$^2$ was obtained by the Berry phase theory of polarization. The Born effective charge ($Z^\star$) was estimated by the finite difference method from the expression 
\begin{equation}
\delta P=\frac{e}{\Omega}Z^\star\cdot\delta u.
\end{equation}
Here, $\delta u$ is the first-order change of the positions of Te and $\Omega$ is the volume of the unit cell. We estimate the $Z^\star$ is about $9.58\ e$ using the data where $u$ is below 0.1 \AA. This is a large value at the same scale of other ferroelectric compounds indicating the strong interactions of the orbitals between Te and Ge along the [111]-direction.   
\par
\begin{figure}
	\includegraphics[scale=0.75]{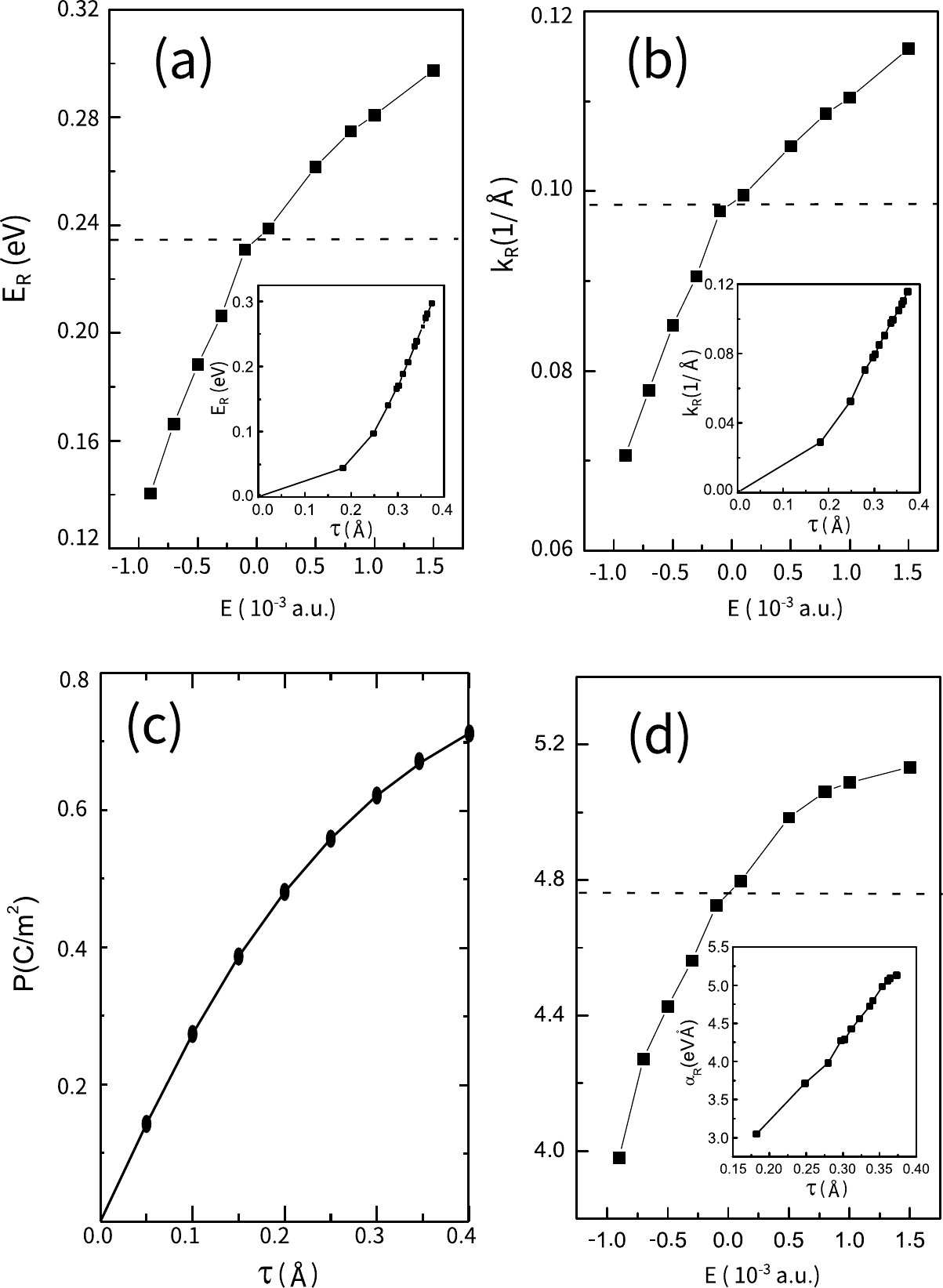}
	\caption{\label{fig:para} The variations of the parameters (E$_R$(a), k$_R$(b) and $\alpha_R$(d)) defined in the Rashba model with the electric field and the electric polarizations as the function of the ferroelectric displacements $\tau$ (c).}
\end{figure}

 Electric field is an effective way to tune the properties of ferroelectric materials. Due to the polarization, electric field will induce displacement of the ions and the electronic structure will change accordingly as in Fig. \ref{fig:para}. The nonlinearity and asymmetry is apparent when the electric field gets larger in the two different directions. The Zener breakdown field $E_{Zb}$ is estimated to be $E_{Zb}=E_g/L\sim0.01 a.u.=51$ MV/cm, where $E_g$ is the band gap and $L$ is the length of the unit cell in the direction of the electric field. In our calculations, the maximum field strength is about one order smaller than the breakdown field in order to avoid instability of the calculation. As expected when the field is along the direction of polarization, the Rashba splitting $E_R$ follows a linear behavior of the electric field ($\vec{D}$) as understood by the Bychkov-Rashba Hamiltonian
\begin{equation}
H_{R}=\alpha_{R}(|\vec{D}|) \vec{\sigma} \cdot\left(\vec{k}_{ \|} \times \vec{e}_{z}\right)
\end{equation}
where $\vec{k}_{ \|}=\left(k_{x}, k_{y}, 0\right)$ and $\vec{e}_{z}=(0,0,1)$. The nature of the phenomenon allows us to use the free two-dimensional electron gases(2DEG) model to get the related parameters. The spin degeneracy of the 2DEG is lifted and the energy dispersion has the form
\begin{equation}
E_{R S O}=\frac{\hbar^{2}}{2 m^{*}}\left(\left|\vec{k}_{ \|}\right| \pm k_R\right)^{2}-E_{R}.
\end{equation}
The Rashba parameter $\alpha_R$, which is related to $E_R$ and $k_R$ by $\alpha_R=2E_R/k_R$, is proportional to the field $\vec{D}$. From our GGA results, the $E_R,k_R$ and $\alpha_R$ are estimated to be 0.24 eV, 0.10 \AA$^{-1}$ and 4.8 eV \AA, respectively, when the field is not applied. The Rashba parameter $\alpha_R$ is in agreement with the experiments \cite{Krempasky2016} and previous theoretical calculations \cite{Sante2012}.

\subsection{SHC and its dependence on the ferroelectric polarization}\label{sec:shc}
\par Since the intrinsic SHC is determined fully by the band structure, it is compatible with the symmetry of the Hamiltonian. We therefore use the symmetry analysis code\cite{Zelezny2017} to figure out the nonzero matrix elements in order to simplify the computation of the third-order tensors. Due to the symmetry constraints, there are only four nonzero elements $\delta_i$($i=0,1,2,\text{ and } 3$). The full SHC tensors are listed as following:
\begin{equation}
\sigma^x:\left(
  \begin{array}{ccc}
    \delta_{0} & 0 & 0 \\
    0 & -\delta_{0} & -\delta_{1} \\
    0 & -\delta_{2} & 0 \\
  \end{array}
\right),
\sigma^y:\left(
  \begin{array}{ccc}
    0 &-\delta_{0} & \delta_{1} \\
    -\delta_{0} & 0 & 0 \\
    \delta_{2} & 0 & 0 \\
  \end{array}
\right),\textrm{ and }
\sigma^z:\left(
  \begin{array}{ccc}
    0 & -\delta_{3} & 0 \\
    \delta_{3} & 0 & 0 \\
    0 & 0 & 0 \\
  \end{array}
\right)
\end{equation}

\begin{figure}
	\includegraphics[scale=0.45]{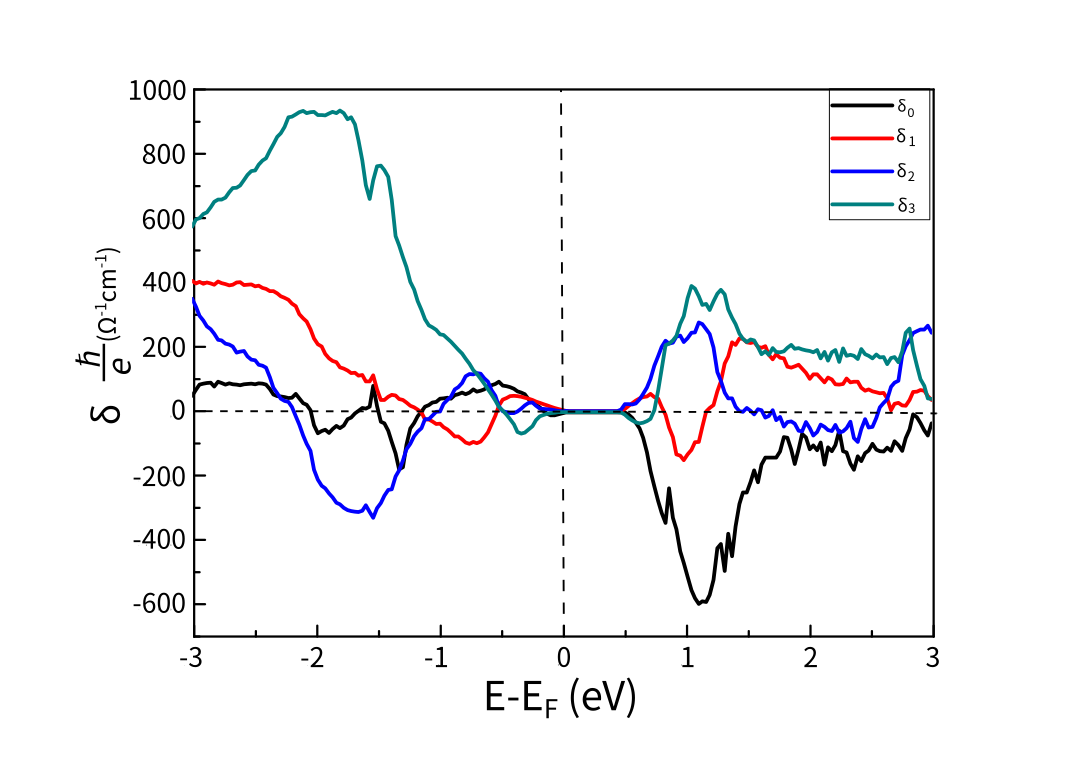}
	\caption{\label{fig:shc_left} The Spin Hall conductivity as a function of the Fermi level.}
\end{figure}

The non-zero elements are calculated at the stationary state with ferroelectric polarization are shown in Fig. \ref{fig:shc_left}. All the tensor components are of the same order of magnitude about 100 $\hbar/e$($\Omega^{-1} $cm$^{-1}$) around the conduction band minimum (CBM) and valence band maximum (VBM). Interestingly, the charge current $I$ and spin polarization $J_s$ are not necessary perpendicular to each other. There are sizable contributions from the $\delta_0$, namely $\sigma^x_{xx(yy)}$ and $\sigma^y_{xy(yx)}$, which means that we can get longitudinal spin Hall signal when the spin polarization is in plane and perpendicular to the electric polarization. This value is peaked when the Fermi energy is about 1.1 eV above the CBM. This feature gives us more freedom to design spintronic devices. The $\sigma^z_{xy}$ is negative around the CBM and VBM. Recent experiments by Rinaldi et al. \cite{Rinaldi2016} reveals that the charge current produced by the spin current in GeTe is opposite
to that found in Pt. As we know, the SHC calculated in Pt\cite{Guo2008} at the Fermi level is 2200 $\hbar/e$($\Omega^{-1} $cm$^{-1}$), which is in good agreement with the experiments\cite{Sinova2015}. Thus the sign of SHC is in agreement with the experiment, however, its magnitude has not been measured yet in GeTe.

\par 
The SHC as a function of the Fermi level is as shown in Fig. \ref{fig:shc_left}. The SHC has the same signs for the electron and hole doping as seen in the figure. The maximum at the VBM and CBM are about -100 $\hbar/e$($\Omega^{-1} $cm$^{-1}$) which are mainly due to the Rashba effect, in the sense that it is sensitively dependent on the electric polarization as can be seen in the following discussions. We see there is much larger SHC amount to 900  $\hbar/e$($\Omega^{-1} $cm$^{-1}$) around 2 eV below the VBM. When comparing with the bandstructure in Fig. \ref{fig:band}, we see that there is a large SOC shift ($\Delta_{SOC}$) of about 0.7 eV of the bands within this energy range. We may conclude that this large value comes from the large SOC induced splitting of the band. 
\par
\begin{figure}
	\includegraphics[scale=0.45]{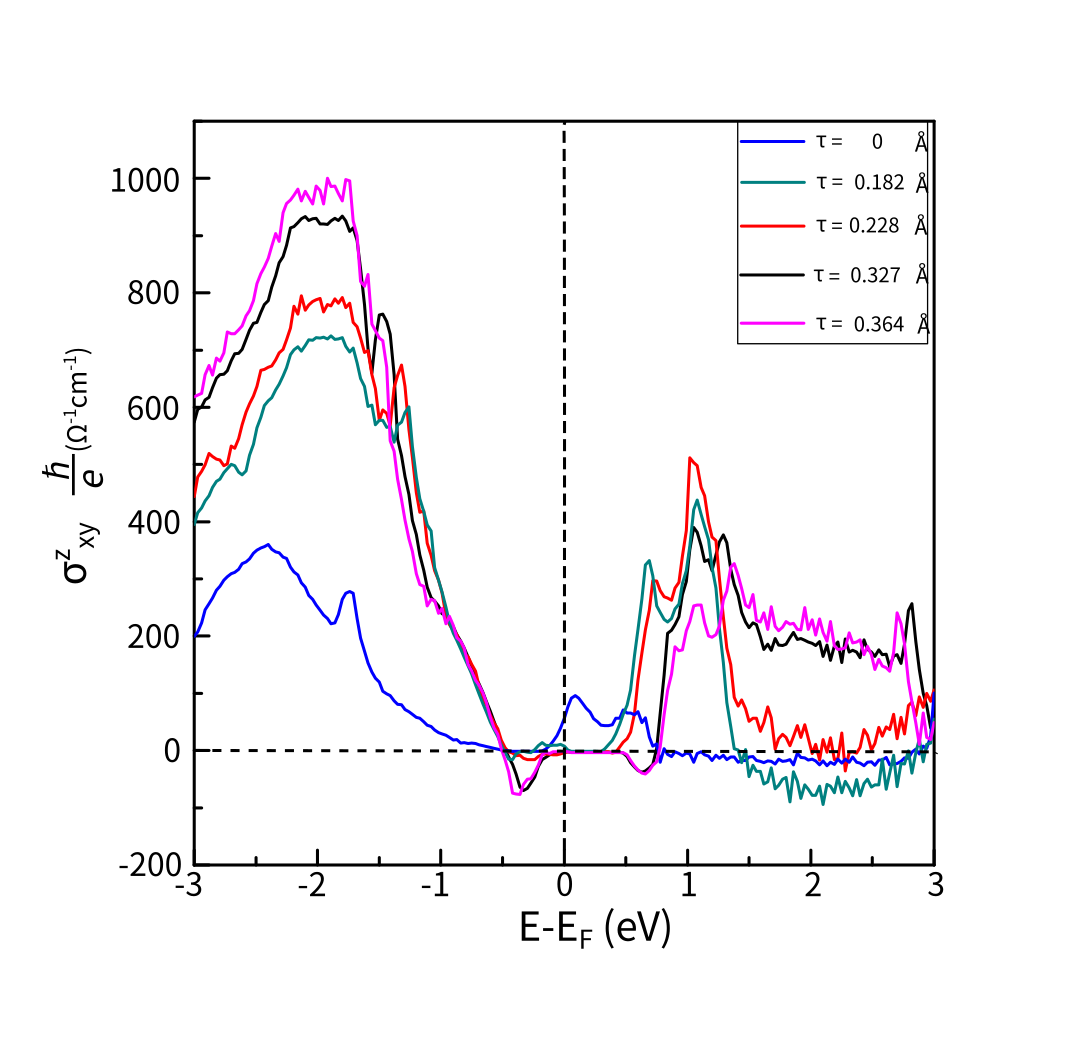}
	\caption{\label{fig:shc-tau} The spin Hall conductivity $\sigma_{xy}^z$ as the function of Te displacement ($\tau$).}
\end{figure}

 
The SHC changes with the polarization as can be seen in Fig. \ref{fig:shc-tau} where the SHC is evaluated at several ferroelectric displacement ($\tau$). At the non-polarized state, GeTe is metal. And its SHC around the Fermi level is about 100 $\hbar/e$($\Omega^{-1} $cm$^{-1}$). The value is peaked amount to 110 $\hbar/e$($\Omega^{-1} $cm$^{-1}$) at 0.2 eV above the Fermi level. The even larger values can be obtained at the energy around -2.0 eV below the Fermi level.  We owe the SHC to the spin Berry curvature of the SOC band due to the equivalent SOC field because there is no Rashba effect. 
When the polarization is increased, we noticed that the SHC around the CBM decreases, and even changes to the negative value. We may propose that the Rashba effect leads to a negative SHC around the CBM and this value increases with the increase of the polarization. The increase of the SHC can be understood from the fact the Rashba constant increases with the electric field as shown already in Fig. \ref{fig:para}. As can be seen from Equ.(2) and (5), $\sigma_{xy}^z$ is an odd function of $z$ and antisymmetric with respect to interchange of $x$ and $y$. In this case, although the reversal of the polarization leads to the reversal of the spin vector winding, it does not lead to sign reversal of the SHC tensor. We have numerically checked the SHC values when the polarization is reversed and find that it remains when the electric polarization is reversed. 

\section{Conclusions}
 To summarize, our theoretical calculation shows that the spin texture is closely related to the ferroelectric polarization. When the ferroelectric polarization is reversed, the spin winding direction is reversed accordingly. This is due to the change of the equivalent coordinate system upon polarization reversal. It should be quite general in FERSC as shown in several compounds. The electric fields are shown to tune the related Rashba parameters. The Rashba parameters increase with the electric field which naturally comes from its origin. The SHCs have contributions from the intrinsic SOC and Rashba effects while the latter can be easily tuned by the external electric field. From the results we suggest that the electric field is can be utilized to moderate the spin and electron transportation properties of FERSC materials. It provides a new dimension to design spintronic devices.

\section{Acknowledgments}
Financial support from National Key R\&D Program of China (No.2017YFB0406403) was greatly acknowledged.

\end{document}